\documentclass[aps,prl,reprint,groupedaddress,showpacs]{revtex4-1}
\usepackage{graphicx}
\graphicspath{{figs/}}

\begin{document}

\title{Spatial Dependence of Electron Interactions in Carbon Nanotubes}

\author{Nicholas Bronn} \altaffiliation{Present Address: IBM T.J. Watson Research Center, 1101 Kitchawan Rd, Yorktown Heights, NY 10598, USA}
\author{Nadya Mason} \email[email: ]{nadya@illinois.edu}

\affiliation{Department of Physics and Materials Research Laboratory \\
University of Illinois at Urbana-Champaign, Urbana, Illinois 61801, USA}

\date{\today}

\begin{abstract}
We report measurements of the spatial dependence of the electron energy distribution in carbon nanotubes, from which electron interactions are determined. Using nonequilibrium tunneling spectroscopy with multiple superconducting probes, we characterize electron transport as ballistic or diffusive, and interactions as elastic or inelastic. We find that transport in nanotubes is generally diffusive, caused by elastic scattering from a few defects. However, local inelastic scattering can be tuned `on' or `off' with a gate voltage.
\end{abstract}


\maketitle

Single-walled carbon nanotubes are often hailed as having ballistic conduction \cite{Tans1997,White1998b}, which enables useful technological applications such as field-effect transistors \cite{Javey2003}, interconnects \cite{Kong1998a}, and qubits \cite{Pei2012a}. Ballistic transport in nanotubes is evident in experiments such as those showing high end-to-end conductance \cite{Mann2003} and Fabry-Perot interference \cite{Liang2001}. However, it is also known that nanotube transport is strongly affected by scattering from defects, as demonstrated, for example, by the modulation of nanotube resistance by tuning single defects `on' (strongly scattering) and `off' (transparent) with a local gate voltage \cite{Bachtold2000a,Bockrath2001} and the saturation of the mean free path at low temperature \cite{Purewal2007a}. A typical 1 $\mu$m long nanotube contains $\sim 3-6$ defects \cite{Bockrath2001,Fan2005}; it is thus unclear from standard transport measurements to what extent such a system can be considered ballistic versus diffusive. In addition, it is difficult to determine via end-to-end transport whether any scattering is inelastic, which affects the electron energy relaxation times relevant for quantum devices \cite{Cleuziou2006b,Herrmann2010}. In general the nature of electron interactions and scattering in 1D systems such as nanotubes is a topic of ongoing interest \cite{NgoDinh2010,Bena2010,Barak2010,Santavicca2010,Karzig2010,Micklitz2011,Ristivojevic2013}. In this Letter, we use nonequilibrium tunneling spectroscopy with multiple superconducting probes to measure the spatial variation of the electron energy distribution along carbon nanotubes \cite{Pothier1997,Chen2009}. This technique allows us to determine whether transport is diffusive or ballistic, as well as if electron scattering is elastic or inelastic. We show that transport in carbon nanotubes can be largely diffusive, via elastic scattering from a few defects. We also show that, although inelastic electron-electron scattering is negligible, inelastic scattering can be induced by gate-tuning defects.

\begin{figure}[b]
	\includegraphics[width=3.375in]{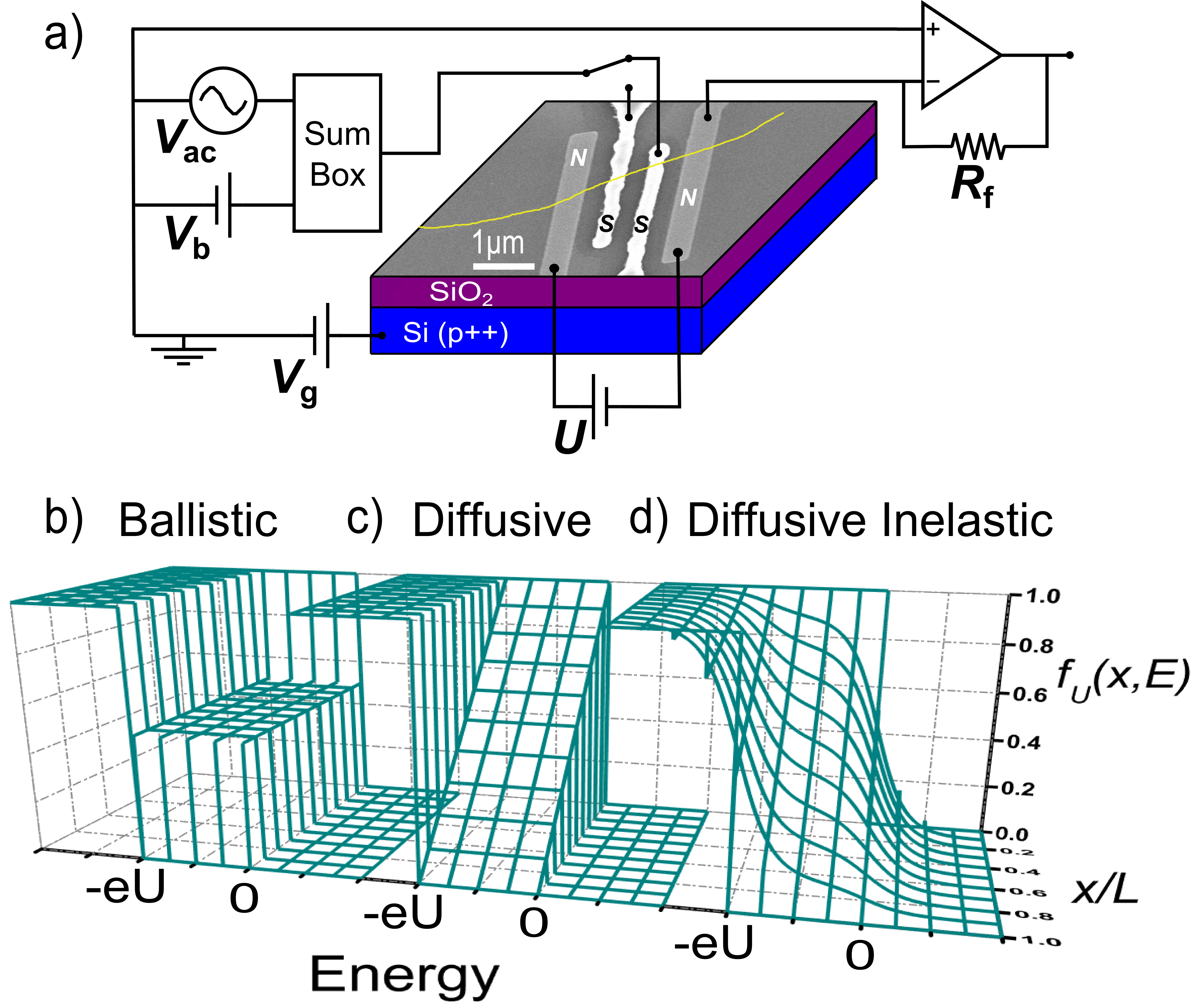}
	\caption{a) Scanning electron micrograph of a device and schematic of the measurement setup (see text). The nanotube, traced in yellow, is contacted by Pd/Au normal (N) end leads (light gray) and superconducting (S) Pb/In tunnel probes (white). Model electron energy distribution functions $f_U(x,E)$ for a nanotube of length $L$ in the b) ballistic, c) diffusive, and d) diffusive with inelastic scattering regimes. The $f_U(x,E)$ in d) are obtained by convolution of the distributions in c) with a thermal broadening function, to simulate the thermalization of electrons due to inelastic electron-electron scattering. These are offset so that each $E=0$ corresponds to the Fermi energy. \label{fig1}}
\end{figure}

\begin{figure}[t]
	\includegraphics[width=3.375in]{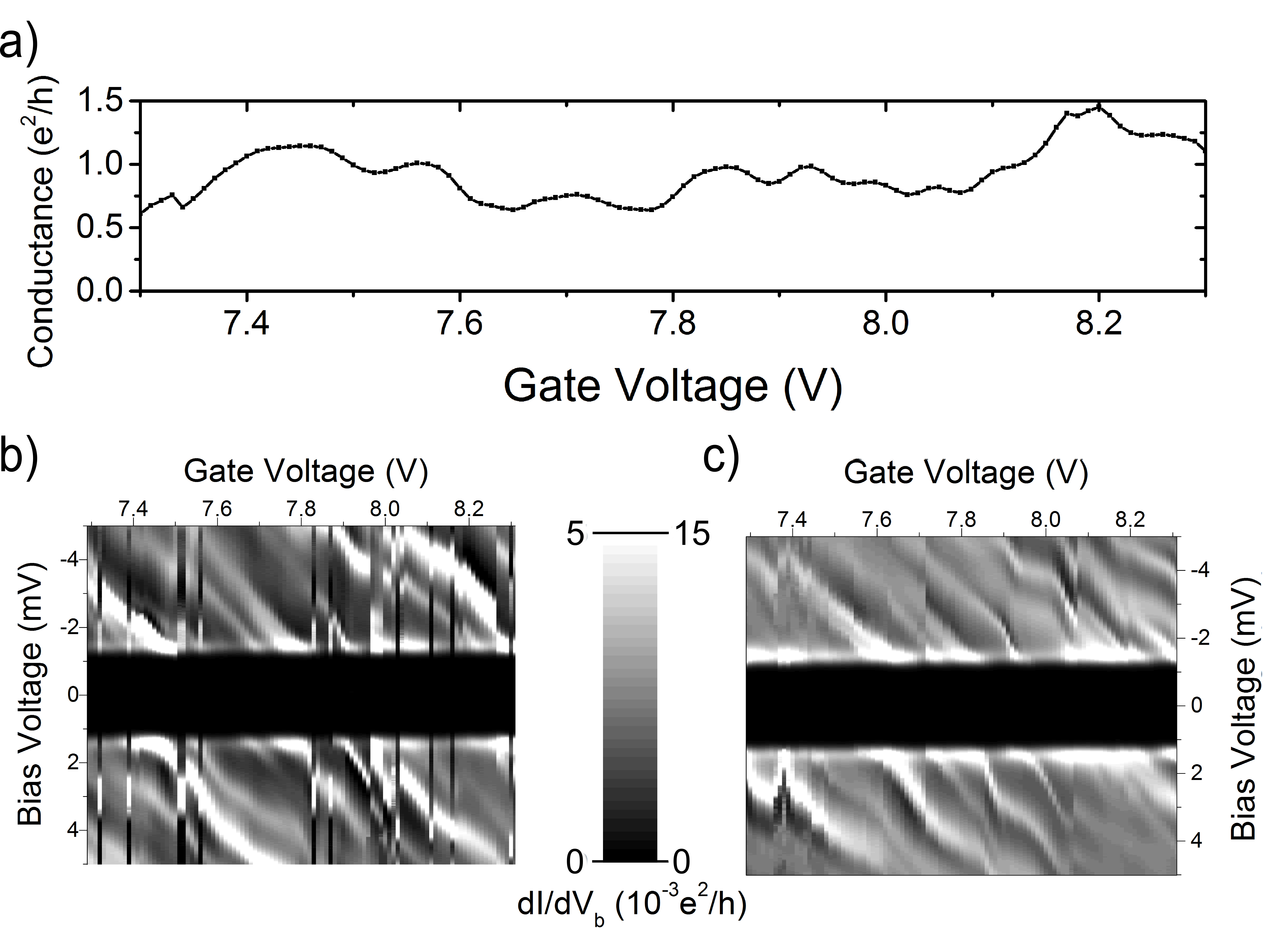}
	\caption{Characterization of nanotube device A at $T = 240$mK (the data for device B are similar). a) End-to-end conductance as a function of gate voltage over a range of one volt. Differential conductance maps with black corresponding to no conduction and white corresponding to maximum conduction, as a function of gate and bias voltage for probe-to-end measurement configurations containing b) no intervening probe and c) an intervening probe. The similarities between the plots demonstrates that an intervening tunnel probe does not add additional features, i.e., it is non-invasive. The difference in conductance scales reflects the difference in tunnel resistance between the probes. Note the superconducting gap with $2\Delta = 2.6$meV is evident for all gate voltages. The intermittent switches in the maps correspond to times when there are charge traps in the thermal oxide later, modifying the nanotube conductance. \label{newfig2}}
\end{figure}

The carbon nanotubes used in our devices are grown from patterned Fe catalysts (of thickness 2 nm) by chemical vapor deposition (CVD) on a degenerately-doped Si substrate having 1 $\mu$m of thermally-grown oxide acting as a gate dielectric, and the devices are fabricated using standard nanofabrication techniques. Here, we discuss detailed measurements of two devices, A and B. Nanotube devices, contacted by two normal metal leads (5 nm Pd/20 nm Au) at a separation of 1.5 $\mu$m, are initially characterized by transport and atomic force microscopy. Devices consisting of single-walled carbon nanotubes (diameters 1-2.5 nm) with high conductance (in this case metallic) are selected for further fabrication and measurement. Next, two superconducting probes (200 nm Pb/30 nm In) of width 200 nm, positioned at approximately one third and two thirds the length of the nanotube (see Fig. \ref{fig1}a), are patterned and deposited via thermal evaporation. Tunnel barriers between the nanotubes and superconducting probes form by oxidation of the Pb in air over the course of several days \cite{Li2013a}, so that the room temperature tunnel resistance $R_\mathrm{T} \sim 2-10$ M$\Omega$ is about 100 times that of the end-to-end resistance $R_\mathrm{end-end} \sim 25-40$ k$\Omega$, which ensures the probing current does not affect the measured distribution function \cite{Gueron1997}. The experiments were performed in a He-3 cryostat at the base temperature of 240 mK, unless otherwise indicated. The devices were characterized at base temperature to ensure the end contacts were well-coupled to the nanotubes, and that the probe deposition did not damage the nanotube, as shown in Fig. \ref{newfig2}. Fig. \ref{newfig2}a shows that end-to-end conductance remains high ($\sim e^2/h$) and has only smooth modulations across an appreciable voltage range. These oscillations are consistent with Fabry-Perot oscillations previously observed in high-conductance carbon nanotubes \cite{Liang2001}. The lack of irregular quantum dot-like features or pinched-off conductance indicate that the nanotube is not damaged by the probe deposition and subsequent oxidation. This is further demonstrated in Figs. \ref{newfig2}b-c, which show conductance maps measured from probe-to-end in configurations without and with an intervening probe, respectively. The conductance map of Fig. \ref{newfig2}c, with an intervening probe, is very similar to that of Fig. \ref{newfig2}b; they both show broad dispersive features that can be associated with Fabry-Perot resonances between the end contacts (the asymmetry in conductance between end contact and tunnel probes makes only one direction of the resonance appear). There is no evidence of additional resonances, localized states or Coulomb blockade due to defects when the conductance is measured in a configuration with an intervening probe, implying that the probes are largely noninvasive. 

For the measurement to determine electron interactions, the nanotubes are driven into steady-state nonequilibrium by application of a voltage $U$ across the ends, then probed by placing the sum of a dc bias voltage $V_\mathrm{b}$ and the ac output of a lock-in amplifier $V_\mathrm{ac}$ on a superconducting tunnel probe, as shown in Fig. \ref{fig1}a. The differential conductance is measured using a lock-in amplifier. The distribution function $f_{\mathrm{nt},U}(E)$ is extracted from the differential tunneling conductance \cite{Pothier1997,Chen2009} by deconvolution of
\begin{eqnarray*}
\bigg(\frac{dI}{dV}\bigg)_U(V_\mathrm{b}) = \frac{1}{R_\mathrm{T}} \int & & \frac{\partial n_{\rm BCS}}{\partial E}(E) n_{\rm nt}(E-eV_\mathrm{b}) \cdot \\
 & & [f_{{\rm nt},U}(E-eV_\mathrm{b}) - f(E)]\,dE
\end{eqnarray*}
derived from the expression for tunnel current \cite{Ingold1992a} with $n_\mathrm{BCS}(E) = \mathrm{Re}\{|E|/\sqrt{E^2-\Delta^2}\}$ the normalized Bardeen-Cooper-Schrieffer (BCS) density of states (DoS) \cite{Tinkham1996} of Pb with a superconducting gap energy of $2\Delta \approx 2.6$ meV, $n_\mathrm{nt}$ the normalized DoS of the nanotube, $f(E) = [\exp((E-E_\mathrm{F})/k_BT)+1]^{-1}$ the Fermi function, and $R_\mathrm{T}$ the room temperature resistance across the tunnel junction.   While $f_{\mathrm{nt},U}$ is simply the Fermi function in the equilibrium case ($U=0$ mV), the evolution of $f_{\mathrm{nt},U}(x,E)$ with $U > 0$ mV along the nanotube gives information about electron relaxation processes and thus electron scattering and interactions \cite{Pothier1997a}. Contacts to the ends of the nanotube at positions $x=0$ and $x=L$ are assumed to be reservoirs of charge carriers in equilibrium \cite{Pothier1997,Chen2009}, and so determine the boundary conditions of the electron energy distribution functions, $f_{\mathrm{nt},U}(0,E) = f(E)$ and $f_{\mathrm{nt},U}(L,E) = f(E+eU)$. Model distribution functions $f_{\mathrm{nt},U}(x,E)$ along a nanotube at $T = 0$ K are depicted in Figs. \ref{fig1}b-d in the following transport regimes: ballistic (no scattering), diffusive (uniformly-spaced elastic scatterers), and diffusive including inelastic electron-electron interactions. In the ballistic regime, every electron (hole) retains its energy as it traverses the nanotube, so the distribution function is the average of those of the end contacts, $f_{\mathrm{nt},U}(x,E) = (1/2)(f(E) + f(E+eU))$; this is a double-step function having a step height of one half at every point along the tube, as shown in Fig. \ref{fig1}b. Elastic scattering tends to localize electrons (holes) closer to the end contact of origination, so that in the limit of a uniform continuum of elastic interactions the distribution function evolves linearly with distance from the end contacts, $f_{\mathrm{nt},U}(x,E) = (1-x/L) f(E) + (x/L) f(E+eU)$; this describes a double-step function whose height varies linearly along the position of the nanotube, as shown in Fig. \ref{fig1}c. Inelastic scattering introduces energy exchange and thus smearing of the distribution function; if this scattering is uniform in the nanotube, as would be the case for intrinsic electron-electron interactions, the smearing is also uniform and should occur for all distribution functions, as depicted in Fig. \ref{fig1}d. Therefore, the shape and spatial variation of the observed distribution functions can determine different transport and scattering regimes. 

\begin{figure}[t]
	\includegraphics[width=3.375in]{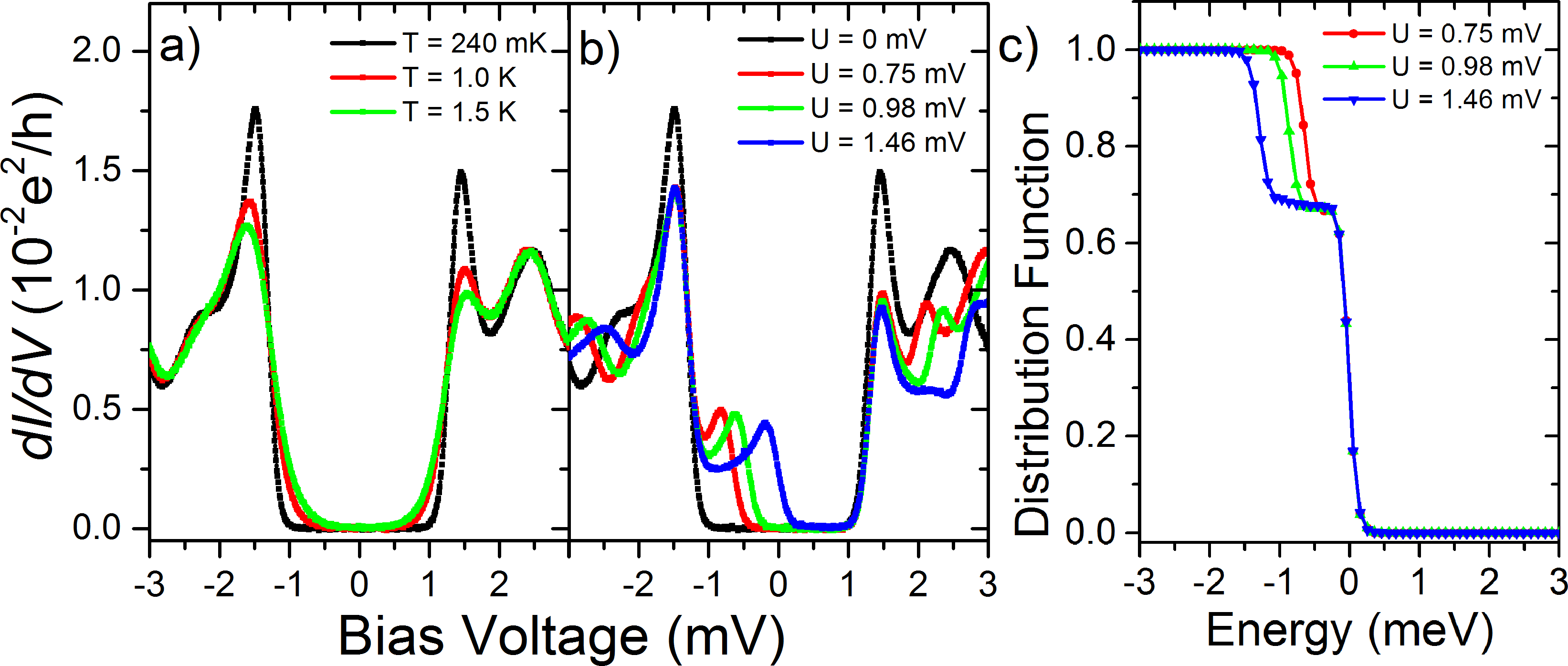}
	\caption{Superconducting tunnel spectroscopy for device A at $V_\mathrm{g} = -5.79$ V. a) Tunneling differential conductance $dI/dV$ vs bias voltage $V_\mathrm{b}$ for the nanotube in equilibrium ($U=0$ mV) at different temperatures, showing that the gap edge peaks become sharper with decreasing temperature, as expected. b) $dI/dV$ vs $V_\mathrm{b}$  for multiple values of bias $U$ at $T=240$ mK, showing `peak-splitting' behavior. c) Electron energy distribution function $f_{\mathrm{nt},U}(E)$ calculated from the data in b), showing a double-step distribution. \label{fig2}}
\end{figure} 
 
Figure \ref{fig2} shows the differential tunneling conductance, $dI/dV$, measured between one of the superconducting probes and one of the end contacts while the bias voltage $V_\mathrm{b}$ is swept. The measurements are performed at gate voltages $V_\mathrm{g}$ for which  the nanotube is in the `open quantum dot' regime, i.e., where the thermal energy $k_\mathrm{B}T$ is smaller than both the quantum level spacing $hv_\mathrm{F}/L$ and Coulomb charging energy $e^2/2C$, yet the conductance is high ($\sim e^2/h$) and does not pinch off to zero because the end contacts are well-coupled to the nanotube \cite{Chen2009}. Figure \ref{fig2}a shows $dI/dV$ for a nanotube in equilibrium ($U=0$ mV), where large, nearly symmetric peaks are evident at the superconducting gap edge, as expected from the BCS DoS. The peaks become sharper at lower temperatures, also as expected. Additional broadened peaks above and below the gap edge can be understood as tunneling peaks through the open quantum dot formed by the nanotube end contacts \cite{Chen2009}. Figure \ref{fig2}b shows $dI/dV$ for a nanotube biased into steady-state nonequilibrium by different voltages $U$. The double-peak structure arises from the convolution of the gap edges with a double-step distribution function; because the second step occurs at an energy of $-eU$ (Fig. \ref{fig1}c), the additional peak in $dI/dV$ is shifted to the right of the gap edge by a value of $U$. Energy distribution functions are calculated from $dI/dV$ using the gradient method of steepest descent (see Ref. \cite{Chen2009}); the calculated, double-step distribution functions corresponding to the data in Fig. \ref{fig2}b are shown in Fig. \ref{fig2}c.

\begin{figure}[t]
	\includegraphics[width=3.375in]{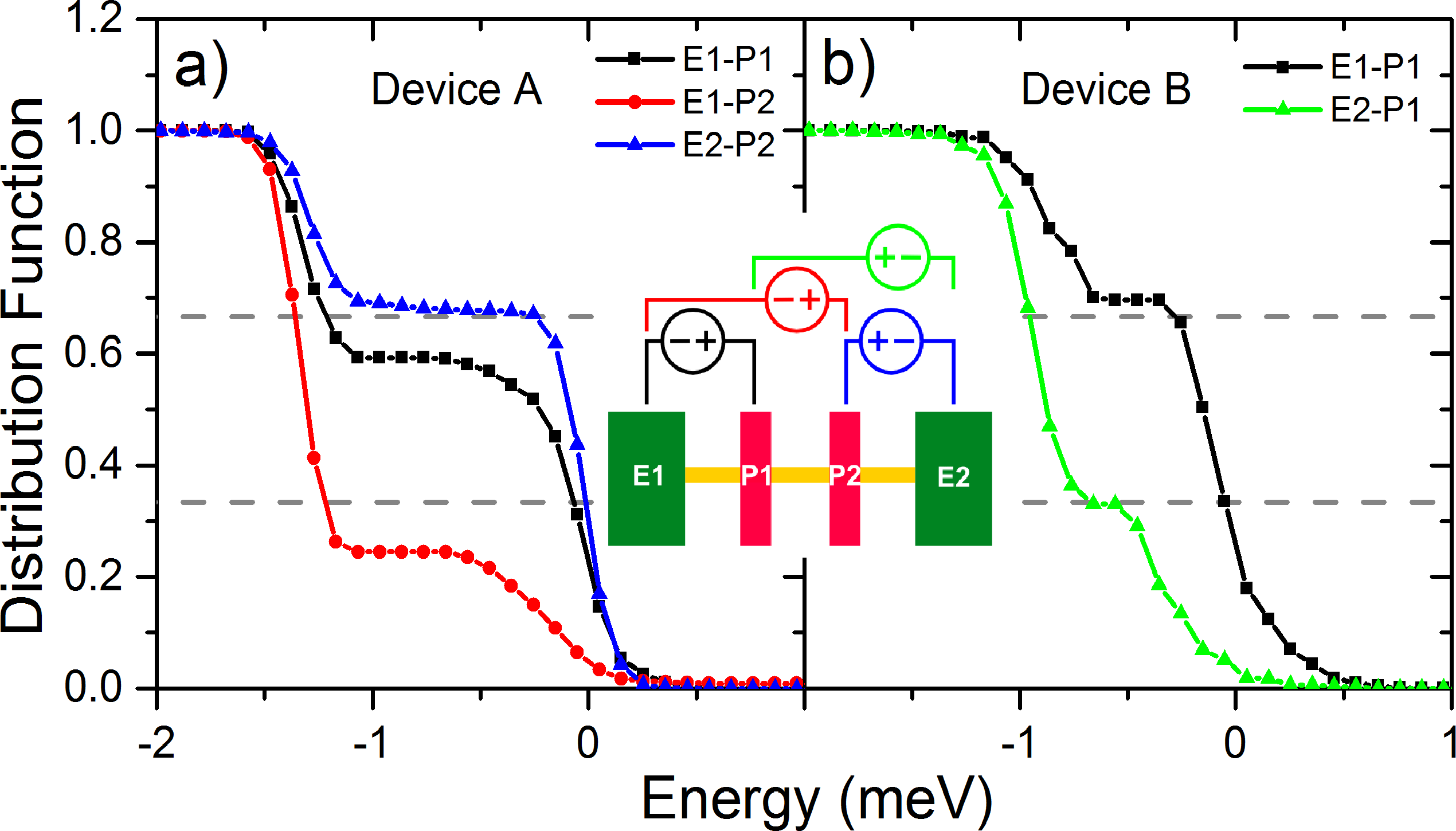}
	\caption{Distribution functions for various measurement configurations for a) device A, at $V_\mathrm{g} = -5.79$ V and $U = 1.48$ mV and b) device B, at $V_\mathrm{g} = 2.74$ V and $U = 0.98$ mV. Step heights of $\sim 1/3$ and $2/3$ (dashed gray lines) occur for probes positioned at $\sim 1/3$ and $2/3$ the distance to the high-bias end of the nanotube, consistent with predictions of diffusive transport. The double step functions indicate an absence of inelastic scattering. Inset: schematic of various probe-to-end measurement configurations. The nonequilibrium voltage $U$ is applied such that it shares `lo' with the measurement. \label{fig3}}
\end{figure}

\begin{figure}[t]
	\includegraphics[width=3.375in]{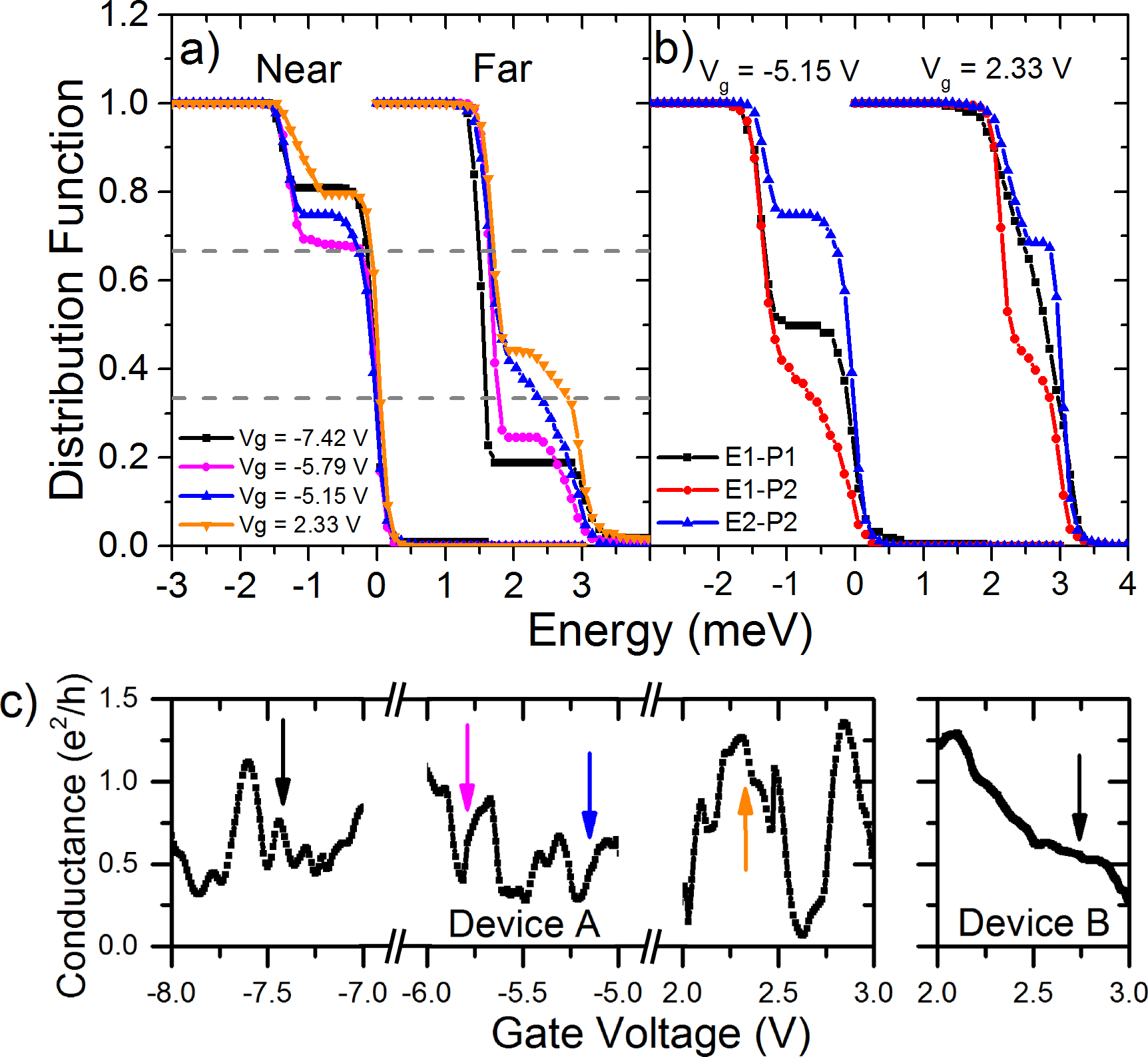}
	\caption{a) Distribution functions for device A for `Near' (End2-Probe2) and `Far' (End1-Probe2) measurement configurations, at multiple gate voltages and at $U=1.46$ mV. The gray dashed lines are at 1/3 and 2/3. The step heights can vary, and occasionally broaden, with gate voltage. b) Distribution functions for various measurement configurations at the same gate voltage. $U=1.46$ and $U=0.98$ mV for the left and right curves, respectively. Smearing of $f_{\mathrm{nt}}(E)$ for E1-P2 at $V_\mathrm{g} = -5.15$ V and for E1-P1 and E1-P2 for $V_\mathrm{g} = 2.33$ V indicates gate-tunable inelastic scattering from defects between P1-P2 and E1-P1, respectively (see text). c) Conductance vs gate voltage for devices A and B, with arrows indicating the $V_\mathrm{g}$ at which the distribution functions in Figs. 3 and 4a,b were measured. \label{fig4}}
\end{figure}

We now discuss electron interactions in the nanotube, determined by studying the spatial and gate voltage dependence of the distribution function. The superconducting tunnel probes lie approximately one third and two thirds along the length of the nanotube (from the grounded normal metal end contact), referred to as the `near' and `far' configurations, respectively. Figure \ref{fig3} shows double-step distribution functions of varying height found for the different near and far configurations, in both devices A and B. The step heights for near configurations have heights of $\sim 2/3$, while those for the far have heights of $\sim 1/3$, independent of the specific device or probe-to-end configuration. The linear dependence of the step heights on position is in good agreement with the model distribution functions for diffusive transport (Fig. \ref{fig1}c), and differs from the expected distributions for ballistic transport (Fig. \ref{fig1}b). Additionally, the sharpness of the steps indicates there is little inelastic electron scattering. Fig. \ref{fig4}a shows the distribution functions for a near and far configuration at multiple gate voltages, demonstrating clustering near the expected values of 2/3 and 1/3 for diffusive transport. However, the step heights do vary with gate voltage, implying that electron energy distribution can be tuned. It has been shown that structural defects occur in CVD-grown metallic nanotubes at a typical separation of $\sim 150$ nm, and that variations in scattering probabilities allows the back-gate voltage to tune individual defects `on' on `off' \cite{Bachtold2000a,Bockrath2001}. Thus, the gate-tunability of the spatial distribution of electron energies suggests that diffusive transport arises from elastic scattering off a few gate-tunable defects. The variation in scattering strength from these defects alters the step height of the distribution functions.

Most of the distribution functions we measured were double-step-like, indicating elastic scattering. However, smeared distributions, indicating inelastic scattering, were also observed at some gate voltages (see blue and orange curves in Fig. \ref{fig4}a), and occurred for all values of $U$. Surprisingly, the inelastic scattering could be tuned `on' and `off' with gate voltage, suggesting that it is caused by defects. This can be seen more clearly in Fig. \ref{fig4}b, which shows distribution functions at fixed gate voltages, but in different measurement configurations, exhibiting both step-like and smeared behavior. In particular, for $V_\mathrm{g} = -5.15$ V, no inelastic scattering is observed between End1 and Probe1 (E1-P1), or End2 and Probe2 (E2-P2), but smearing consistent with inelastic scattering is apparent between End1 and Probe2 (E1-P2). This implies that the inelastic scattering takes place locally, between Probe1 and Probe2. A similar analysis for $V_\mathrm{g} = 2.33$ V shows that inelastic scattering occurs between End1 and Probe1 at that gate voltage. Inelastic scattering caused by intrinsic electron-electron interactions or via nanotube phonon modes should not be locally gate-tunable in this way \cite{Pothier1997,Chen2009}. Rather, these results imply that individual defects are being turned `on' to cause inelastic electron scattering in nanotubes. It is possible that such scattering is inelastic because the defects are well-coupled to external baths, such as phonon modes in the substrate \cite{Perebeinos2009}. We note that no intrinsic effect of electron-electron inelastic scattering was observed, despite predictions of strong electron interactions in a 1D system \cite{NgoDinh2010,Bena2010,Karzig2010,Micklitz2011}.

Finally, we show that the interaction regimes inferred from the electron energy distribution functions cannot be determined via standard end-to-end transport measurements. Fig. \ref{fig4}c shows nanotube end-to-end conductance as a function of gate voltage in regions around which the nonequilibrium experiments are performed. Broad resonances are evident in the conductance, consistent with the Fabry-Perot-like oscillations seen in Fig. 2a. There are no trends in the data that indicate the nature of the transport or the amount of inelastic scattering. In fact, the gate voltages at which the most inelastic scattering occurs are $V_\mathrm{g} = -5.15$ and $2.33$ V, which feature the highest ($1.23e^2/h$) and lowest ($0.45e^2/h$) end-to-end conductance, respectively, while conductance at the gate voltages in which inelastic scattering is absent ($V_\mathrm{g} = -7.42$ and $-5.79$ V) fall between those values (device B exhibited much more inelastic scattering). This is consistent with previous work demonstrating that the existence of defects in carbon nanotubes is difficult to observe from end-to-end conductance alone \cite{Bachtold2000a,Bockrath2001,Purewal2007a}.

Nonequilibrium tunneling spectroscopy offers new insight into transport in carbon nanotubes. In particular, these experiments indicate that electron transport in nanotubes can be diffusive, so that the ideal ballistic transport cannot be assumed. In addition, while elastic scattering from structural defects is well-known, these results indicate that inelastic scattering from other types of defects must also be considered. Gate-tunable inelastic scatterers may appear in similar surface-supported nanostructures such as graphene, which is relevant to designing nanoscale quantum devices requiring long energy relaxation times.

We thank Norman O. Birge, Matthew J. Gilbert, and Alex Levchenko for useful discussions. This work was supported by the NSF under grant DMR-0906521 and was carried out in part in the Frederick Seitz Materials Research Laboratory Central Facilities, University of Illinois.

\bibliography{CNTpaper}

\end{document}